\documentclass{PoS}
\pdfoutput=1

\usepackage{multirow}
\usepackage{graphicx}
\usepackage{grffile}
\usepackage{epstopdf}
\usepackage{amsmath}
\usepackage{placeins}
\usepackage{tikz}
\usetikzlibrary{arrows}
\usetikzlibrary{patterns}
\usetikzlibrary{plotmarks}
\def \taille{0.8}

\newcommand{\mr}{\mathrm}
\newcommand{\fm}{\,\mr{fm}}
\newcommand{\MeV}{\,\mr{MeV}}
\newcommand{\GeV}{\,\mr{GeV}}

\newcommand{\lsim}{ {\
\lower-1.2pt\vbox{\hbox{\rlap{$<$}\lower5pt\vbox{\hbox{$\sim$}}}}\ } }
\newcommand{\gsim}{ {\
\lower-1.2pt\vbox{\hbox{\rlap{$>$}\lower5pt\vbox{\hbox{$\sim$}}}}\ } }

\title{Finite-volume corrections to the leading-order hadronic contribution to $g_\mu-2$}

\ShortTitle{FV corrections to $a_\mu^{HVP,LO}$}

\author{\begin{minipage}{\textwidth}
\begin{center}
\speaker{Rehan Malak}$^{1,2}$, Zoltan Fodor$^{3,4,5}$, Christian Hoelbling$^3$,
Laurent Lellouch$^{1}$, Alfonso Sastre$^3$, Kalman Szabo$^4$\\[0.4cm]
for the \textbf{ Budapest-Marseille-Wuppertal collaboration}\\[0.2cm]
{\it
$^1$ CNRS, Aix Marseille U., U. de Toulon, CPT, UMR 7332, F-13288 Marseille,
France\\
$^2$ CNRS, CEA, Maison de la Simulation,
USR 3441, F-91191 Gif-sur-Yvette Cedex, France\\
$^3$Department of Physics, Bergische Universit\"{a}t Wuppertal, Gaussstr. 20, 
D-42119 Wuppertal, Germany\\
$^4$IAS/JSC, Forschungszentrum J\"ulich, D-52425 J\"ulich, Germany\\
$^5$Inst. for Theor. Physics, E\"otv\"os University, P\'azm\'any P. s\'et. 1/A, H-1117
Budapest, Hungary}
\end{center}

 E-mail: \email{rehan.malak@cpt.univ-mrs.fr}
\end{minipage}

}

\abstract{We present preliminary results of a 2+1-flavor study of
  finite-volume effects in the lattice QCD computation of the
  leading-order hadronic contribution to the muon anomalous magnetic
  moment. We also present methods for obtaining directly the
  invariant hadronic polarization function, $\Pi(Q^2)$, and the Adler
  function at all discrete lattice values of $Q^2$, including
  $Q^2=0$. Results are obtained with HEX-smeared clover fermions.}

\FullConference{The 32nd International Symposium on Lattice Field Theory,\\
		23-28 June, 2014\\
		Columbia University New York, NY}

\begin{document}
%\vspace{-0.15in}

\section{Introduction}
%\vspace{-0.4cm}
There still is an unexplained 3.4 $\sigma$ discrepancy between the
experimental measurement \cite{Bennett:2006fi} of the anomalous
magnetic moment of the muon, $a_{\mu}=\frac{(g-2)_{\mu}}{2}$, and the
theoretical determination based on the standard
model \cite{Knecht:2014sea}. The leading uncertainties in the
theoretical computation are associated with two non-perturbative QCD
corrections : the hadronic vacuum polarization contribution
$a_{\mu}^\mr{HVP,LO}$ at $\mathcal{O}(\alpha)^2$ and the hadronic
light-by-light scattering term at $\mathcal{O}(\alpha^3)$. The former
is presently best determined via dispersion relations applied to
experimental cross sections for $e^+e^-$ annihilation and $\tau$
decays into hadrons \cite{Davier:2010nc,Hagiwara:2011af}. In
anticipation of the future Fermilab E989 experiment, whose goal is to
divide by $4$ the error on the measurement of
$a_\mu$ \cite{Venanzoni:2014ixa}, several groups are now working on
computations of $a_{\mu}^{HVP,LO}$ using lattice QCD simulations (see
e.g.\ for their latest
contributions \cite{Aubin:2006xv,Boyle:2011hu,Burger:2014lna,Chakraborty:2014mfa,Francis:2014qta,Golterman:2014rda}). This
approach will provide a valuable
\textit{ab-initio} cross-check for phenomenological determinations.

We present a method for obtaining the scalar
polarization function, $\Pi(Q^2)$, and the Adler function at all
discrete lattice values of $Q^2$ including zero, directly from the
vector-vector correlation function. We then use this approach and the usual
one based on the vacuum polarization tensor, to study finite-volume
effects in the lattice computation of $\Pi(Q^2)$ and $a_\mu^\mr{HVP,LO}$. In particular, we
present preliminary results of a dedicated study of these effects at a
fixed lattice spacing of $0.104\fm$ and pion mass $M_\pi\sim 292\MeV$,
for lattices ranging in spatial size $L$ from $2.5$ to
$8.3\fm$.

\section{Usual and new ways to obtain $a_{\mu}^\mr{HVP,LO}$ on the lattice}
%\vspace{-0.4cm}
Based on a formula first derived in \cite{Lautrup:1971yp}, it was
shown in \cite{Blum:2002ii} how $a_{\mu}^\mr{HVP,LO}$ can be obtained
from the polarization tensor computed directly in Euclidean spacetime,
using lattice QCD simulations. Schematically one computes the Fourier
transform of the expectation value of the product of two
electromagnetic quark currents, $J_\mu$, at Euclidean lattice momenta
$Q$:
\begin{equation}
\Pi_{\mu\nu}(Q)=a^4\sum_x \langle J_\mu(x)J_\nu(0)\rangle e^{iQ\cdot x}
\ .\end{equation}
Neglecting Euclidean $O(4)$ violations, the vacuum polarization
tensor, $\Pi_{\mu\nu}(Q)$, can be written in terms of a single
invariant function $\Pi(Q^2)$, as:
\begin{equation}
\Pi_{\mu\nu}(Q)=(  Q_{\mu}Q_{\nu}-\delta_{\mu\nu}Q^{2} ) \Pi(Q^{2})
\ .\end{equation}
Note that the decomposition assumes that $\Pi_{\mu\nu}(Q=0)=0$, which
is certainly true in infinite volume, and is required for the photon
to remain massless. But it is not necessarily the case in a finite
spacetime. Thus, we distinguish two methods. In the first, which we
call \textit{``usual without subtraction''}, we define
$\Pi(Q^{2})\equiv \Pi_{\mu\nu}(Q)/(
Q_{\mu}Q_{\nu}-\delta_{\mu\nu}Q^{2} )$. In the \textit{usual method
with subtraction}, we take $\Pi(Q^{2})\equiv
[\Pi_{\mu\nu}(Q)-\Pi_{\mu\nu}(0)]/(
Q_{\mu}Q_{\nu}-\delta_{\mu\nu}Q^{2} )$ \cite{LehnerTalk:2014}. Here we consider only the
spatial, $\Pi_{ii}(Q)$, $i=1,2,3$, components of the polarization
tensor. The resulting $\Pi(Q^{2})$ is then fitted as a function of $Q^2$
and extrapolated to $Q^2=0$ to perform the required additive
renormalization, $\hat{\Pi}(Q^{2}) = \Pi(Q^{2}) - \Pi(0)$.  The same
fit is used to integrate the polarization function with a known QED
kernel $w_{\Pi}(Q^2)$, yielding the muon anomalous magnetic moment
through:
\begin{equation}
\label{eq:amuint}
a_{\mu}^\mr{HVP,LO} = 4\pi^2  \Big(\frac{\alpha}{\pi}\Big)^2 \sum_{f=u,d,s,\ldots} q_f^{2} \int_{0}^{\infty} dQ^2 w_{\Pi}(Q^{2}) \hat{\Pi}_f(Q^2)
\ ,\end{equation}
where $q_f$ is the charge of quark flavor $f$ in units of
$e$. $a_{\mu}^\mr{HVP,LO}$ is then studied as a function of
simulations parameters and interpolated and/or extrapolated to the
physical values of the quark masses and to the continuum and infinite
volume limits.

A precise determination of $a_{\mu}^\mr{HVP,LO}$ on the lattice is
particularly challenging because the kernel $w_{\Pi}$ peaks at
$Q^2 \sim (m_{\mu}/2)^2$, which is smaller than the lowest
non-zero momenta $(2\pi/L,T)$ available in current simulations on
$T\times L^3$ lattices with periodic boundary conditions. Moreover, 
$\Pi_{\mu\nu}(Q)$ is noisy for small values of $Q^2$.

Besides the \textit{usual} methods for determining $\Pi(Q^2)$
described above, we consider a third approach, which circumvents the
problem that $\Pi(Q^2=0)$ is not directly accessible. Thus, it
eliminates the systematic and statistical error associated with the
extrapolation to $Q^2=0$ required to renormalize $\Pi(Q^2)$ and to
describe this function for the values of $Q^2$ which contribute most
to $a_{\mu}^\mr{HVP,LO}$.

This second set of methods considers Fourier derivatives of the polarization
tensor:~\footnote{Fourier derivatives are also considered
in \cite{Chakraborty:2014mfa,Francis:2014qta}, for instance, but only
in the time direction and at $Q^2=0$.}
\begin{equation}
\label{eq:JJderivatives}
\partial_\rho\partial_\sigma \Pi_{\mu\nu}(Q)= - a^4 \sum_{x}x_{\rho}x_{\sigma}
\langle J_\mu(x)J_\nu(0)\rangle e^{iQ\cdot x}
 \ .\end{equation}
We call it the -\textit{2nd derivative}- method. By appropriately choosing
the indices and the four-momentum, one can obtain directly the desired
polarization scalar through:
\begin{equation}
\label{eq:PiQ2direct}
\Pi(Q^2)=\partial_\mu\partial_\nu \Pi_{\mu\nu}(Q)\vert_{Q_{\mu}=Q_\nu=0}, \qquad \mu\ne\nu
 \ ,\end{equation}
or through
\begin{equation}
\Pi(Q^2)=-\frac12
\partial_\mu\partial_\mu \Pi_{\nu\nu}(Q)\vert_{Q_{\rho}=0,\rho\ne\nu}, \qquad \mu\ne\nu
 \ .\end{equation}
One can also obtain the Adler function:
\begin{equation}
\mathcal{A}(Q^2)=Q^2\frac{\partial\Pi(Q^2)}{\partial
Q^2}=-\frac12 \partial_\mu\partial_\mu \Pi_{\mu\mu}(Q)\vert_{Q_{\mu}=0}
 \ .\end{equation}
In all cases, one may choose to sum over repeated indices, including spatial
and/or temporal components, depending on the symmetries of the lattice
under consideration.  Here we focus on the polarization scalar and we
consider only results obtained from Eq.~(\ref{eq:PiQ2direct}) with the
spatial components, $\Pi_{ij}(Q)$, $i\ne j=1,2,3$, of the polarization
tensor. In general, the {\em 2nd derivative} methods have the advantage
that they also work to obtain information at $Q^2=0$, thereby guaranteeing that the
interesting values of $Q^2$ are reached by a controlled
interpolation. A possible drawback is that the factor of $x_\rho
x_\sigma$ term, in Eq.~(\ref{eq:JJderivatives}), emphasizes long-distance
contributions which are more noisy and more subject to finite-volume
effects.

In both the {\em usual} and {\em 2nd derivative} approaches we split up
the fit of $\Pi(Q^2)$ vs $Q^2$ into two regions, as suggested
in \cite{Golterman:2014rda}. In the -\textit{low}- $Q^2$ region, we
fit a [1,1], three parameter Pad\'e to the 4 lowest available momentum
points. The -\textit{high}- $Q^2$ results up to $1\GeV^2$ are fit to
another Pad\'e. In that way, the fit to the low region, which
contributes most to $a_{\mu}^\mr{HVP,LO}$, and has larger statistical
errors, is not distorted by the more precise results at higher values
of $Q^2$. The integration yielding
$a_{\mu}^\mr{HVP,LO}$ is split accordingly, with no particular
matching at $0.2\GeV^2$. Note that, in this proceedings, we only
integrate up to $Q^2=1\GeV^2$ and thus consider a quantity
$a_{\mu}^\mr{HVP,LO}(Q^2\le 1\GeV^2)$ which is not quite the HVP
contribution to the muon anomalous magnetic moment.

\section{Finite-volume study}
%\vspace{-0.4cm}
The preliminary finite-volume study presented here is based on four
Budapest-Marseille-Wuppertal, $N_f = 1+1+1+1$ ensembles that
were generated for the recent calculation of the neutron-proton mass
difference in QCD+QED \cite{Borsanyi:2014jba}. Here we consider $N_f =
2+1$ valence flavors that couple only to the $SU(3)$-color
components of the links, with bare masses adjusted to reproduce the
isospin averaged $u$ and $d$ quark and the $s$ quark masses of the $N_f = 1+1+1+1$
calculation. The results are obtained using a tree-level
$\mathcal{O}(a^2)$ improved Symanzik gauge action, together with
tree-level clover-improved Wilson fermions. The gluon fields undergo three
steps of HEX smearing before being coupled to the quarks. The hadronic
vacuum polarization tensor, $\Pi_{\mu\nu}$, is computed with a local
vector current at the source (indexed by $\nu$) and a conserved vector
current at the sink (indexed by $\mu$). For the present study we
neglect disconnected components, which are expected to be small
compared to the connected contributions that we retain, and which
should not appreciably modify the finite-volume behavior.

The four ensembles considered here are those from
\cite{Borsanyi:2014jba} with $\beta=3.2$, corresponding to
$a=0.104\fm$, and with $M_\pi\sim 292\MeV$. The bare mass parameters
used in the valence sector are $am_{ud}=-0.077$ and $am_s=-0.050$.
The four ensembles differ only in their volumes, with the
spatial size of spacetime, $L$, ranging from $2.5$ to $8.3\fm$. The
relevant characteristics of the ensembles are:
\begin{center}
    \begin{tabular}{|c|c|c|c|c|c|c|}
\hline
T/a & L/a & $M_{\pi}$ (MeV) & T (fm) & L (fm) & $M_{\pi}$T & $M_{\pi}$L  \\
\hline
48 & 24 & 295.2(1.4) (0.50\%)               & 5.0 & 2.5 & 7.5 & 3.7 \\
64 & 32 & 292.6(7) (0.23\%)              & 6.7 & 3.3 & 9.9 & 4.9 \\
96 & 48 & 292.0(6) (0.20\%)              & 10.0 & 5.0 & 14.8 & 7.4 \\
64 & 80 & 292.1(3) (0.12\%)              & 6.7 & 8.3 & 9.9 & 12.3 \\
\hline
\end{tabular}

\end{center}
Three of the four simulations have $T/L=2$ and all are on asymetrical
lattices. The pion is light enough to allow the $\rho$ to decay into
two pions in the infinite volume limit. It is important to note that
besides the $u$ and $d$ quarks being more massive than physical, the
strange is not finely tuned to its physical
value here. Thus, one should not expect these quark's contributions
to the polarization scalar and $a_{\mu}^\mr{HVP,LO}$ to take on their
(precise) physical values. This is all the more true that we leave out
the finite renormalization, $Z_V$, of the local electromagnetic quark
current, which contributes only an overall factor to these quantities.

\begin{figure}[t]
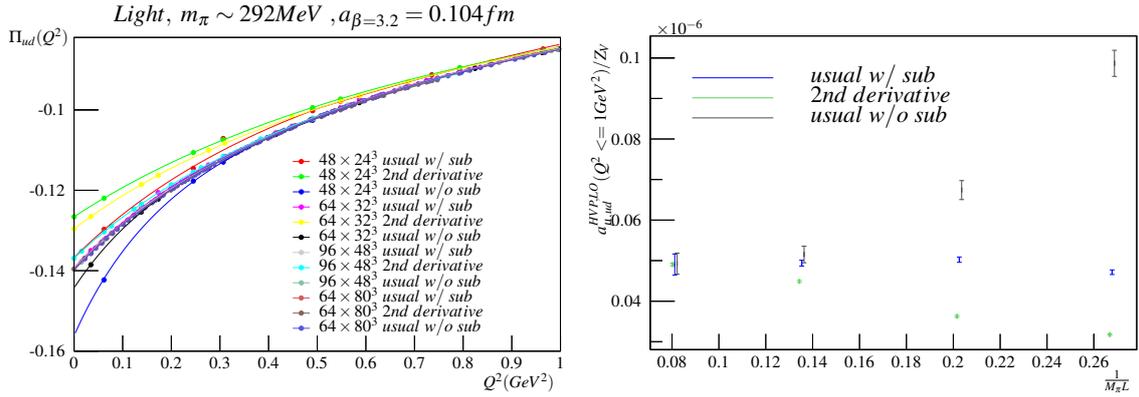

    \begin{minipage}{.5\textwidth}
        \scalebox{\taille}{\input{CVC_0.tex}}
    \end{minipage}
    \begin{minipage}{.5\textwidth}
        \scalebox{\taille}{\input{volume_man_FV3hex_CVC_0.tex}}
    \end{minipage}
\caption{\label{fig:FVlchan}{\em (Left panel)} $\Pi_{ud}(Q^2)$ vs $Q^2$ for $M_\pi\sim 292\MeV$, as obtained using the {\em usual} and {\em 2nd derivative} (Eq.~(\protect\ref{eq:PiQ2direct})) methods in four volumes and with all other lattice parameters fixed. The data points are the values obtained from the current-current correlation function and its Fourier derivatives. The curves are the corresponding fits. {\em (Right panel)} $a_{\mu,ud}^\mr{HVP,LO}(Q^2\le 1\GeV^2)$ vs $1/M_{\pi}L$ obtained from the polarization functions in the left panel.}
\end{figure}

We begin by studying the light, up-down quark contribution to
$a_\mu^\mr{HVP,LO}$. In the left panel of Fig.~\ref{fig:FVlchan} we
show $\Pi_{ud}(Q^2)\equiv \Pi_{u}(Q^2)=\Pi_{d}(Q^2)$ vs $Q^2$ for the
four different volumes, obtained using the \textit{usual} and {\em 2nd
derivative} (Eq.~(\ref{eq:PiQ2direct})) methods. Also shown are the
fits to Pad\'es described above. While for the smallest volume the
three methods yield results which differ significantly at low $Q^2$,
this difference reduces as the volume is increased, and the three
methods give fully compatible results in the largest volume.  This
convergence of the methods in the limit of large volumes is also
clearly visible in the right panel of Fig.~\ref{fig:FVlchan}, where
the values of $a_{\mu, ud}^\mr{HVP,LO}(Q^2\le 1\GeV^2)$, obtained by
integrating the fit functions for $\hat\Pi_{ud}(Q^2)$ according to
Eq.~(\ref{eq:amuint}), are plotted against $1/M_{\pi}L$. We choose
$M_{\pi}L$ because it is dimensionless and because $1/M_\pi$ is the
longest correlation length in the system. Since the dependence of
$M_\pi$ on $L$ is very weak, the $1/M_{\pi}L$ dependence shown here is
equivalent to a $1/L$ dependence.

While results from the three methods converge in the large-volume
limit, in smaller volumes the finite-size corrections are significant
in some cases. In the smallest volume, with $L=2.5\fm$ or
$LM_\pi=3.7$, the finite-volume correction on
$a_{\mu,ud}^\mr{HVP,LO}(Q^2\le 1\GeV^2)$, obtained using the {\em
2nd derivative} method, is $\sim 35\%$. It is even larger for the {\em
usual method without subtraction}: around $200\%$. In the {\em
2nd derivative} case, it is reduced to below $10\%$ by the time $L\gsim
5\fm$. Only results obtained from the {\em usual method with
subtraction} do the finite-volume effects remain small for all volumes
considered.

An interesting feature of the {\em 2nd derivative} method is that it
features significantly smaller statistical errors on
$a_{\mu}^\mr{HVP,LO}(Q^2\le 1\GeV^2)$ than the {\em usual method
without subtraction}. This remain true to a much smaller extent for
the {\em usual method with subtraction}. In the former case, it is
mainly due to the fact that the {\em 2nd derivative} method eliminates
the noisy $\Pi_{\mu\nu}^{ud}(0)$, as does {\em usual method with
subtraction}. The additional statistical improvement compared to the
{\em usual method with subtraction} results from the fact that the
{\em 2nd derivative} method allows the extraction of
$\Pi(Q^2=0)$. This constrains the statistical fluctuations of the
fitted $\Pi(Q^2)$ vs $Q^2$ in the very important low-$Q^2$ region. And
though we do not investigate this issue here, this additional
constraint will also reduce systematic errors by replacing the usual
extrapolation by an interpolation.

\begin{figure}[t]
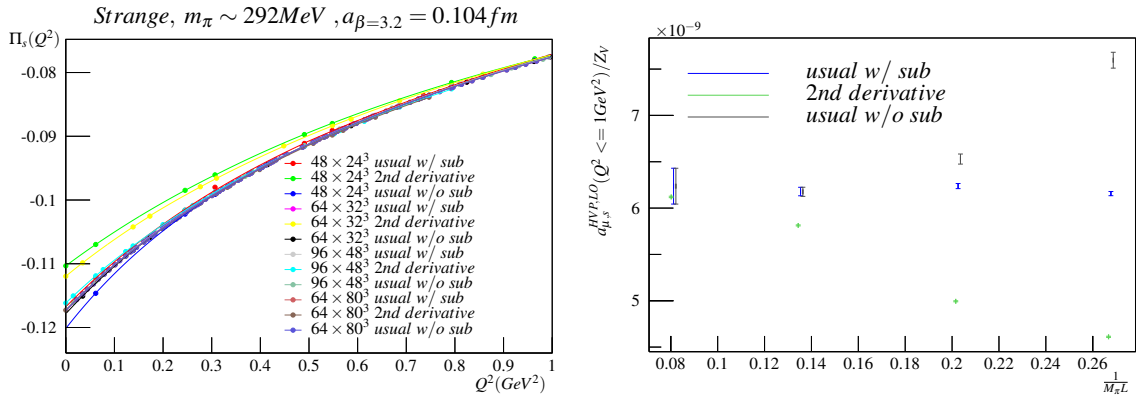

    \begin{minipage}{.5\textwidth}
        \scalebox{\taille}{\input{CVC_1.tex}}
    \end{minipage}
    \begin{minipage}{.5\textwidth}
        \scalebox{\taille}{\input{volume_man_FV3hex_CVC_1.tex}}
    \end{minipage}
\caption{\label{fig:FVschan}Same as Fig.~\protect\ref{fig:FVlchan}, but for the strange-quark contribution.}
\end{figure}

We now turn to the strange-quark contribution to $a_\mu^\mr{HVP,LO}$
and perform the same study of finite-volume effects as for the light
contribution. The corresponding results for $\Pi_s(Q^2)$ vs $Q^2$ and
$a_{\mu,s}^\mr{HVP,LO}(Q^2\le 1\GeV^2)$ vs $1/M_{\pi}L$ are shown in
Fig.~\ref{fig:FVschan}.  For both quantities, the same general
features, as were observed for the light contribution, are seen
here. In particular, the results obtained from the {\em usual method
with subtraction} show no volume dependence for the lattices
considered. On the other hand, significant finite-volume effects are
still observed for the two other methods in smaller volumes, but these
disappear as one goes to larger lattices. They are, nevertheless, much
smaller than in the light case. For the strange contribution, the
finite-volume correction, in the smallest volume with $L=2.5\fm$ or
$LM_\pi=3.7$, is now $\sim 25\%$ on $a_{\mu}^\mr{HVP,LO}(Q^2\le
1\GeV^2)$ obtained using the {\em derivative} method and $\sim 20\%$
when it is obtained using the {\em usual approach without subtraction}. By the time one
reaches $L=5.0\fm$ or $LM_\pi=7.4$, the effect is not statistically
significant for the {\em usual} method and below $5\%$ for the {\em
derivative} approach.

\section{Conclusion}
%\vspace{-0.4cm}
In addition to the {\em usual} methods for obtaining the polarization
scalar, $\Pi(Q^2)$, which consist in dividing the polarization tensor
$\Pi_{\mu\nu}(Q^2)$, or its subtracted counterpart
$[\Pi_{\mu\nu}(Q^2)-\Pi_{\mu\nu}(0)]$, by $(Q_\mu
Q_\nu-\delta_{\mu\nu}Q^2)$, we have considered a {\em 2nd derivative}
method based on Fourier derivatives of quark-electromagnetic-current
two-point functions. This method yields $\Pi(Q^2)$ directly, obviating
the need to divide by $(Q_\mu Q_\nu-\delta_{\mu\nu}Q^2)$, which
clearly is not possible when $Q_\mu=0$. One advantage of this method
is that it gives direct access to $\Pi(0)$.  In addition, it does so
in a way which is consistent with the results obtained at other values
of $Q^2$. Indeed, the study presented above shows that it may be
dangerous to try to combine $\Pi(0)$ obtained through derivatives with
$\Pi(Q^2)$, $Q^2>0$, obtained in the {\em usual} ways, as the methods
have significantly different systematic errors. The study also shows
the advantage of being able to determine $\Pi(0)$ directly: it reduces
the statistical errors on the result for the muon anomalous magnetic
moment, and a similar reduction is anticipated for the systematic
errors.

Using the {\em usual-without-subtraction}, {\em
usual-with-subtraction} and {\em 2nd derivative} methods for determining
$\Pi(Q^2)$, we have conducted a dedicated study of finite-volume
effects. We find that the size of
finite-volume corrections depends strongly on the method used to
obtain $\Pi(Q^2)$, on the quark-contribution considered and, of
course, on volume.

The $u$ and $d$ quarks in this study were chosen to be light enough
for the $\rho$ to be a resonance in infinite volume. Thus we expect
that the physics which governs finite-volume effects here is, at least
qualitatively, similar to that which is at play for light quarks at
their physical mass. If that is the case, for the contributions of the
$u$, $d$ and $s$ quarks, the {\em usual method with subtraction} is
clearly preferable from the point of view of finite volume effects.
As the volume is made larger, the fact that the {\em 2nd derivative}
method leads to smaller statistical and $\Pi(Q^2)$-vs-$Q^2$ fit
uncertainties makes it increasingly attractive. Of course, further
study is required to understand the extent to which these conclusions
carry over to the situation of physical, light-quark masses.

%\vspace{-0.4cm}
\section*{Acknowledgments}
\vspace{-0.4cm}
We thank Eric B. Gregory and Craig McNeile for their contributions in the early stages of the project.
Computations were performed using HPC resources provided by the PRACE
Research Infrastructure Resources JUGENE at FZ J\"ulich and Fermi at
CINECA, with further HPC resources provided by GENCI-IDRIS (Grant
No. 52275) and FZ J\"ulich. This work was supported in part by the
OCEVU Labex (ANR-11-LABX-0060) and the A$^\star$MIDEX project
(ANR-11-IDEX-0001-02), funded by the "Investissements d'Avenir" French
government program and managed by the ANR, and by DFG grants FO 502/2,
SFB-TR 55.

\bibliographystyle{JHEP_notitle}
\bibliography{pos_rmalak_2014,E989_alternatives}
\end{document}